\documentclass[conference]{IEEEtran}
\IEEEoverridecommandlockouts
\usepackage{cite}
\usepackage{amsmath,amssymb,amsfonts}
\usepackage{graphicx}
\usepackage{textcomp}
\usepackage{xcolor}
\usepackage{algorithm}
\usepackage{algpseudocode}
\usepackage{comment}
\algnewcommand{\Initialize}[1]{%
  \State \textbf{Initialize:}
  \Statex \hspace*{\algorithmicindent}\parbox[t]{.8\linewidth}{\raggedright #1}
}

\def\BibTeX{{\rm B\kern-.05em{\sc i\kern-.025em b}\kern-.08em
    T\kern-.1667em\lower.7ex\hbox{E}\kern-.125emX}}

\begin{document}

\title{Distributed Vehicular Dynamic Spectrum Access for Platooning Environments \\
\thanks{Copyright © 2020 IEEE. Personal use is permitted. For any other purposes, permission must be obtained from the IEEE by emailing pubspermissions@ieee.org. This is the author’s version of an article that has been published in the proceedings of 2020 IEEE 91st Vehicular Technology Conference (VTC2020-Spring) and published by IEEE. Changes were made to this version by the publisher prior to publication, the final version of record is available at: http://dx.doi.org/10.1109/VTC2020-Spring48590.2020.9128929. To cite the paper use: P. Sroka, P. Kryszkiewicz, M. Sybis, A. Kliks, K. S. Gill and A. Wyglinski, "Distributed Vehicular Dynamic Spectrum Access for Platooning Environments," 2020 IEEE 91st Vehicular Technology Conference (VTC2020-Spring), 2020, pp. 1-5, doi: 10.1109/VTC2020-Spring48590.2020.9128929. https://ieeexplore.ieee.org/document/9128929. 

The work has been realized within the project no. 2018/29/B/ST7/01241 funded by the National Science Centre in Poland and US NSF 1547296.}
}

\author{\IEEEauthorblockN{Pawe\l~Sroka}
\IEEEauthorblockA{\textit{Poznan University of Technology}\\
Poznan, Poland \\
pawel.sroka@put.poznan.pl}
\and
\IEEEauthorblockN{Pawe\l~Kryszkiewicz}
\IEEEauthorblockA{\textit{Poznan University of Technology}\\
Poznan, Poland \\
pawel.kryszkiewicz@put.poznan.pl}
\and
\IEEEauthorblockN{Micha\l~Sybis}
\IEEEauthorblockA{\textit{Poznan University of Technology}\\
Poznan, Poland \\
michal.sybis@put.poznan.pl}
\and
\IEEEauthorblockN{Adrian~Kliks}
\IEEEauthorblockA{\textit{Poznan University of Technology}\\
Poznan, Poland \\
adrian.kliks@put.poznan.pl}
\and
\IEEEauthorblockN{Kuldeep~S.~Gill}
\IEEEauthorblockA{\textit{Worcester Polytechnic Institute} \\
Worcester, MA, USA \\
ksgill@wpi.edu}
\and
\IEEEauthorblockN{Alexander~Wyglinski}
\IEEEauthorblockA{\textit{Worcester Polytechnic Institute} \\
Worcester, MA, USA \\
alexw@wpi.edu}

}

\maketitle

\begin{abstract}
In this paper, we propose a distributed Vehicular Dynamic Spectrum Access (VDSA) framework for vehicles operating in platoon formations. Given the potential for significant congestion in licensed frequency bands for vehicular applications such as 5.9 GHz. Our approach proposes to offload part of the intra-platoon data traffic to spectral white-spaces in order to enhance vehicular connectivity in support of on-road operations. To enable VDSA, a  Bumblebee-based decision making process is employed which is based on the behavioral models of animals, is employed to provide a means of distributed transmission band selection. Simulation results show the distributed VDSA framework improves the leader packets reception ratio by 5\%, thus indicating its potential to increase in reliability of intra-platoon communications.
\end{abstract}

\begin{IEEEkeywords}
V2X Communications, System Coexistence, Dynamic Frequency Selection, Vehicular Dynamic Spectrum Access
\end{IEEEkeywords}

\section{Introduction}
A coordinated movement of a formations of vehicles, known as \textit{platooning}, is one prospective application of the emerging autonomous driving technology, where a group of self-driving cars and/or trucks forms a convoy, that is led by a lead vehicle \cite{LPT+16}. For safety operations of autonomous driving, \textit{e.g.}~using the Cooperative Adaptive Cruise Control (CACC) \cite{Raj2000}, wireless communications between vehicles can be used to exchange control information within the platoon.

From the perspective of data exchange within the vehicular network, such as within a platoon formation, it can be realize it using short-range wireless communication schemes, such as Dedicated Short-Range Communications (DSRC). However, various studies have show that solutions based on the IEEE 802.11p and Wireless Access in Vehicular Environment (WAVE) standards are susceptible to medium congestion when the number of communicating cars is large \cite{REDDY18,VUK2018} . An alternative approach to remedy this issue is to offload traffic to other frequency bands, such as underutilized television channels (known as TV White Spaces (TVWS) \cite{Harrison_2010,Beek_2012}); this  concept is referred to as Vehicular Dynamic Spectrum Access (VDSA) \cite{Chen_2011,Chen_2012}. Fixed locations and relatively stable transmission parameters of Digital Terrestrial Television (DTT) transmitters provide attractive spectrum opportunities in TVWS that can be reused for vehicular communications. 
In order to employ the VDSA framework for Vehicle-to-Vehicle (V2V) networks, the allocation of channels to the primary users (\textit{i.e.} DTT transmitters) and the associated power levels for the areas of interest should be known. This can be achieved using infrastructure support such as dedicated databases, although, in some areas this might be unavailable. An~alternative approach is to apply VDSA in a distributed manner, where each vehicle or group of vehicles (\textit{e.g.} platoon) selects a transmission channel based on spectrum sensing. This approach has been proposed in \cite{8347136,8690978,8417762}, which relies on the behavioral model of bumblebee foragers to provide efficient channel sensing and selection.

In this paper, we investigate the idea of partial intra-platoon traffic offloading from the Control Channel (CCH) in congested 5.9 GHz band to TVWS using the VDSA framework. To enable channel sensing and selection We adopt the Bumblebee-based algorithm from \cite{8347136,8690978,8417762} to dynamically select the transmission channel in TVWS in a distributed manner. We assume that all platoon vehicle perform spectrum sensing and share the results with the platoon leader,which is responsible for the selection of the transmission band. Furthermore, we evaluate the proposed distributed VDSA framework using computer simulations aided with realistic DTT signal power obtained with the measurements described in \cite{VTC_DSA}.

The rest of this paper is organized as follows: in Section {II}, we introduce the system model used in this work. Section {III} presents the several aspects invoking DTT signal power and interference modelling. Section {IV} describes the details of the adapted Bumblebee-based decision making process to the platooning scenario. Section {V} presents the achieved simulation results with applied distributed VDSA, and Section {VI} concludes the paper.

\section{System Model}
\begin{figure}[htbp]
\centerline{\includegraphics[width=0.5\textwidth]{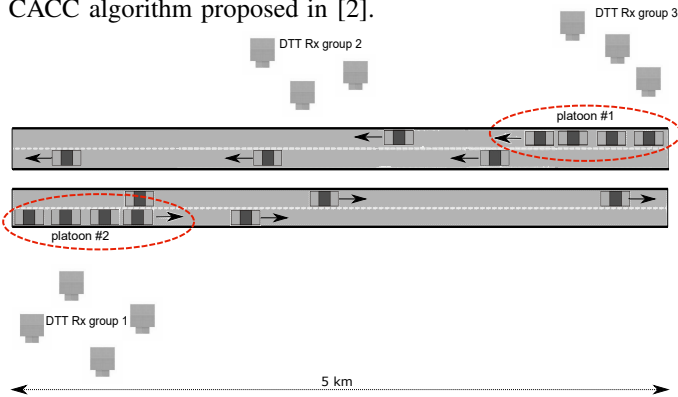}}
\caption{Investigated 4-lane motorway scenario with two platoons (marked with red ellipses) and the DTT receivers' locations (building symbols on both sides of the motorway). The platoons  move in opposite directions bypassing each other in the middle section of the motorway.}
\label{fig_scenario}
\end{figure}

In this work. we consider a motorway scenario such as the one presented in Fig. \ref{fig_scenario}, where multiple platoons of cars travel among other vehicles. The platooning vehicles are assumed to be autonomous with their mobility controlled using the CACC algorithm proposed in \cite{Raj2000}.\\

\vspace{-11mm}

To obtain a straightforward analysis and without loss of generality, we consider only two platoons moving in the opposite directions and located on the outer lanes of the 4-lane motorway. The behavior of each platoon is affected by a preceding car that periodically decreases its velocity from 130 km/h to 100 km/h and then accelerates back to 130 km/h with a period equal to 30 s. On the inner lanes of the motorway, additional non-platooning vehicles are randomly deployed with a density of 20 cars/km/lane.

We assume that all vehicles are using wireless communications based on the IEEE 802.11p standard \cite{IEEE80211} to disseminate Cooperative Awareness Messages (CAMs) that contain information on selected mobility-related parameters. Wireless communications is also employed to facilitate platooning using CACC, with every vehicle receiving the required mobility information from the preceding car and from the platoon leader. Communications between non-platoon vehicles using CAMs, which is transmitted with a frequency of 10 Hz, is realized in the CCH at 5.9 GHz. For platooning vehicles that transmit CAMs with a frequency of 5 Hz, additional dedicated CACC messages are sent using the TVWS frequency bands. The CACC transmission interval is set to 200 ms, such that the combined CAM+CACC messaging frequency is equal to 10 Hz. Simultaneous operations in two different frequency bands is achieved with the use of dual-band transceivers.

In this work, we employ a VDSA algorithm described in Section \ref{sec:bumblebee} in order to support the intra-platoon communications such that the Signal-to-Interference-and-Noise Ratio (SINR) of the V2V transmission is maximized. Given the distributed approach under investigation, the selection of transmission frequency in the TVWS band will rely on the sensing of the bands under consideration. Energy sensing has been employed for all DTT channels considered for reuse, with the procedure of sensing the particular bands distributed across all platoon vehicles. We assume the results of the channel sensing operation are then transmitted to the platoon leader, which is responsible for selecting the band for intra-platoon transmission.

To investigate the impact of VDSA in TVWS  we consider the presence of DTT receivers near the motorway that need to be protected. These are located in longitudinal positions and distances to motorway, as indicated in Table \ref{tab_DTTpos}.
\begin{table}[htbp]
\caption{DTT receivers' positions used in the considered scenario as the protected locations of the primary users.}
\label{tab_DTTpos}
\begin{tabular}{|c|c|c|c|}
\hline
 DTT Rx id & long. position [m] & dist. to motorway [m] & Group \\
 \hline
 1 & 240 & 120 & 1 \\
 \hline
 2 & 4520 & 164 & 3 \\
 \hline
 3 & 4320 & 244 & 3 \\
 \hline
 4 & 320 & 45 & 1 \\
 \hline
 5 & 1687 & 80 & 2 \\
 \hline
 6 & 4112 & 304 & 3 \\
 \hline
 7 & 632 & 140 & 1 \\
 \hline
 8 & 485 & 270 & 1 \\
 \hline 
 9 & 1463 & 154 & 2 \\
 \hline
 10 & 2087 & 127 & 2 \\
 \hline
\end{tabular}
\end{table}
 
\section{DTT signal power modeling}
\label{sec:DTT}
\begin{figure}[htbp]
\centerline{\includegraphics[width=0.5\textwidth]{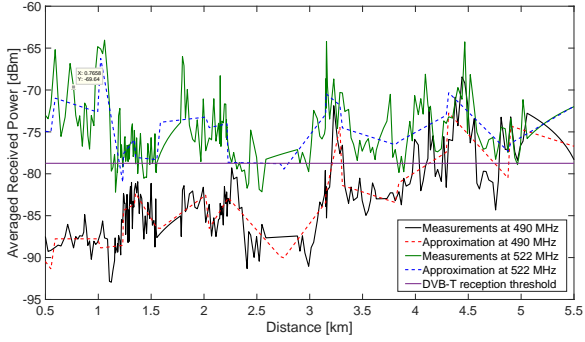}}
\caption{Received DVB-T signal power along the route at channel 23 and 27; solid horizontal line represents the minimum required DVB-T reception threshold.}
\label{fig_rec_power}
\end{figure}

In order to reliably evaluate the proposed VDSA spectrum allocation mechanism, a test data set is required. A DTT received power measurement campaign was conducted for a TV band (470 MHz - 692~MHz), with the details of the measurements described in \cite{VTC_DSA}. The resulting models of measured Digital Video Broadcasting - Terrestrial (DVB-T) signal power in two channels (490 MHz and 522MHz) versus the route distance are shown in Fig.~\ref{fig_rec_power}. Additionally, the DVB-T reception threshold of -78.75 dBm was determined to be the minimum DTT received power allowing for successful TV program reception in a thermal noise-limited environment, as explained in \cite{VTC_DSA}. Only these two channels contained DTT signal exceeding test receiver noise floor.  
In order to reduce the burden of storing the raw measurement results, as well as to capture possible variations caused by shadowing, we divided the measured samples into segments depending on location. For each segment, linear approximations using a~least squares approach are employed. The resulting curves are shown in Fig.~\ref{fig_rec_power}. In the simulator, the sensed DTT received power is modeled using these linear functions along with a log-normal shadowing random variable added. The standard deviation of shadowing was estimated for each measurements segment separately, with its value varying from 1.5~dB to 4.5~dB.

Even though the considered band contains non-negligible DTT signals only in two DTT channels, their influence on the adjacent band should be modeled considering the out-of-band DTT transmitters leakage (modeled typically by Adjacent Channel Leakage Ratio - ACLR) and imperfect DSRC receiver selectivity (modeled typically by Adjacent Channel Selectivity - ACS). These two effects combined, describing the transmitter-receiver interference coupling, are characterized by an Adjacent Channel Interference Ratio (ACIR) that is characteristic for a given frequency offset between both systems. Therefore, in the scenario under consideration, the effective power of the DTT signals, \textit{e.g.}, observed by Bumblebee algorithm energy detector, in each of the considered DTT channels is calculated as a sum of the power values of the DTT signals in a given location scaled by the proper ACIR values. Modeling of the ACIR between DSRC and DVB-T systems has been described in \cite{VTC_DSA}.  

\section{Distributed VDSA based on bumblebee behavioral model}
\label{sec:bumblebee}

Given the highly stochastic time-varying channel environment associated with vehicular communications  our proposed distributed VDSA framework employs a Bumblebee-based VDSA algorithm~\cite{8347136,8690978,8417762} for the purpose of independent decision-making such that it can enable vehicle-to-vehicle (V2V) communication links with minimal latency and be robust to rapidly changing operating conditions. In this paper, the Bumblebee-based algorithm is integrated with a realistic IEEE 802.11p simulator and evaluated with respect to its feasibility as a distributed VDSA mechanism for platooning support.

\begin{algorithm}[!ht]
\caption{Bumblebee-based VDSA Algorithm}
\label{algo}
\begin{algorithmic}[1]
\Procedure {BumblebeeDSAAlgorithm}{$Ch$, $\mathbf{T}$, $C$, $S$}
\Initialize{Set the \textbf{Radio:1} channel to the initial DTT band $Ch$}
\vspace{0.4em}
\State \textbf{Channel selection interval} ($1000$ ms):  
\For{each 200 ms interval}
\State \textbf{Sensing interval} (150 ms):
\State Select the sensed DTT channel for each vehicle
\State Sense the selected DTT bands
\State \textbf{Transmission interval} (50 ms):
\State Switch to current transmission channel $Ch$
\State Perform intra-platoon communication and sensing in $Ch$ for 50 ms
\EndFor
\For{each sensed DTT channel $i$}
\State Calculate the average sensed energy $E_i$ in $i$
\If {$E_{Ch} > (E_{i} - C)  \And   E_{i} \leq \mathbf{T}$ } 
\State Switch to new channel: $Ch \leftarrow i$
\Else 
\State Keep using same channel $Ch$
\EndIf
\EndFor
\State Configure \textbf{Radio:1} with selected channel $Ch$
\EndProcedure
\end{algorithmic}
\end{algorithm} 

Algorithm~\ref{algo} describes the Bumblebee VDSA algorithm employed in a dual-transceiver IEEE 802.11p communications network. In this work, we have setup the VDSA framework within the DTT band since the primary users of this band are relatively stable when compared to other wireless frequency bands~\cite{srikanthdsa}. In this framework \texttt{Radio:0} is constantly tuned to the 5.9~GHz CCH channel such that it can transmit and receive CAMs, while \texttt{Radio:1} is working in the DTT band with channel switching employed. We initialized the VDSA model by assigning an arbitrarily selected DTT channel to \texttt{Radio:1}. The~implemented algorithm then periodically selects every second the DTT channel that is the best suited for intra-platoon transmission in terms of the highest expected SINR. Channel selection is based on energy levels detected during the sensing phases. As \texttt{Radio:1} needs to switch between sensing and transmission, this is performed in a cyclical manner, with each 200~ms cycle comprising of a 150~ms sensing period and a 50~ms transmission period. This duty cycle duration was selected in order to comply with the CACC messaging frequency. It should be noted that doing the sensing interval all DTT channels can be sensed by different platoon vehicles, while during the transmission phase only the sensing of currently used transmission channel is possible.

Once the energy values $E$ for all the DTT channels are measured in 5 consecutive cycles, we select the best channel that is characterized by the lowest energy level if the energy does not exceed the threshold value $\mathbf{T}$. An additional cost $C$ is also introduced to reduce the unnecessary channel switching as these can adversely affect the performance due to the need to reconfigure the transceiver. Eq.~\ref{switchDecision} describes the switching decision, where $\bar{E_{Ch}}$ is the mean signal energy for the current channel, and $\bar{E_{i}}$ is the mean energy value of the $i$th channel in the DTT band.

\vspace{-5mm}

\begin{align}
\begin{tabular}{c}
Switching\\ Decision
  \end{tabular}
= \left\{
  \begin{tabular}{l}
 $ \bar{E_{Ch}} \leq (\bar{E_{i}} - C)$, \text{``Stay"} \\
   Otherwise ,\hspace{18pt} \text{``Switch" }
  \end{tabular}
\right.
\label{switchDecision}
\end{align}

Once the new channel is selected, channel switching can be triggered, \textit{i.e.}, \texttt{Radio:1} can switch to the new channel in the DTT band, where the highest SINR is expected. Following this, the procedure is repeated in the next channel selection period. 

\section{Simulation Experiments}

The performance of VDSA in TVWS frequencies for intra-platoon communications has been evaluated using system-level simulations of the considered scenario. We used a simulation tool developed in C++, that was calibrated and used also in work described in \cite{VUK2018,Syb19}. The duration of a single simulation represented a platoon traveling the distance of 5 km over a time period of 140~s.
The distributed VDSA framework was applied with 5 adjacent DTT channels included with center frequencies located at 490 MHz, 498 MHz, 506 MHz, 514 MHz, and 522 MHz. Since the IEEE 802.11p standard uses 10 MHz bandwidth and the center frequencies for VDSA transmission are assumed to be equal to the DTT channel frequencies, the signals from neighbouring bands may partially overlap. Moreover, the two edge bands, \textit{i.e.}, at 490 MHz and 522 MHz, are used by DTT transmitters, hence strong inter-system interference is expected, which is described in Section \ref{sec:DTT}

We compared the performance of a Bumblebee-based VDSA framework with the reference system that used transmissions operating only in the 5.9~GHz band and a system utilizing also transmissions in TVWS frequencies but without channel switching (a~fixed channel at 506 MHz was used). For the Bumblebee-based approaches, three switching cost values were considered: 0~dB (no cost), 3 dB, and 6 dB. One should also note that for the TVWS band, the default IEEE 802.11p CSMA sensitivity threshold was increased by 10 dB in order to reduce the impact of transceiver blocking by high DTT power levels.

\begin{figure}[htbp]
\centerline{\includegraphics[width=0.5\textwidth]{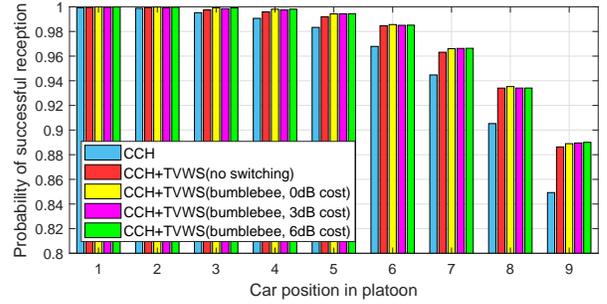}}
\caption{Probability of successful reception of leaders' packets.}
\label{fig_leaderP}
\end{figure}

One important performance indicator for intra-platoon communications is transmission reliability, which can be represented by the ratio of successful receptions to the total number of messages transmitted. It is assumed this indicator should be kept over 99\% to enable safe autonomous platooning. Fulfilling of such requirement, especially in the case of high density traffic on motorways, might not be possible when only CCH is used  \cite{VUK2018}. This is reflected also with the results shown in Fig.~\ref{fig_leaderP}, which presents the successful reception rate of packets transmitted by the platoon leader versus the vehicle position in the platoon. Employing intra-platoon communications in TVWS frequencies with additional CACC packets improves the reception rate, especially for vehicles at the tail of the platoon, what is mainly due to the lower messaging rate of CCH and the higher transmission range in the TVWS. However, the reception probability is almost the same for all transmission configurations using TVWS. The application of the Bubmlebee-based algorithm does improve slightly the reception rate for the vehicles at the tail of the platoon, but the observed difference is marginal. Surprisingly, the best reception rate is observed with the additional cost of 6~dB applied in switching procedure. Hence, one may conclude that high sensitivity to changes in the observed energy results in the selection of poor frequency bands.

\begin{figure}[htbp]
\centerline{\includegraphics[width=0.5\textwidth]{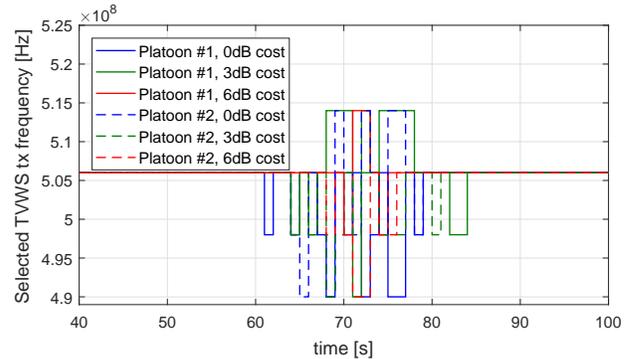}}
\caption{Example of frequency switching for two platoons moving in opposite directions and different considered switching costs $C$. It can be observed that frequency band switching occurs when the platoons are close to each other in the time interval of 60-80~s. Frequency bands at 490~MHz and 522~MHz are rarely or never selected due to the presence of DVB-T transmissions.}
\label{fig_switchEx}
\end{figure}

Besides the reliability requirement, another factor indicating the performance of VDSA is the introduced cost with respect to the need to change the operating frequency for the entire platoon, which requires additional lend of dissemination and coordination effort. Such cost depends on the number of frequency switches performed by each platoon. In this work, we have considered three configurations with different expected rate of switching depending on the selected value of the $C$ parameter in the Bumblebee-based algorithm, with the example of selected frequencies for different configurations shown in Fig.~\ref{fig_switchEx}. It can be observed that when the platoons are sufficiently separated, for both of them the middle frequency band (\textit{i.e.}, 506~MHz) is selected since the interference from the DTT is the lowest. However, when the platoons are closing, their transmission in the same band starts to affect the channel selection procedure and switching starts. Since no coordination between platoons is applied, it is observed that sometimes both platoons select the same band (the one with the lowest energy), which does not improve the performance. It indicates that an additional factor (\textit{e.g.} randomness) could be introduced to improve the channel selection procedure.

\vspace{-5mm}

The average number of switches, representing the expected additional cost of VDSA, is given in Table \ref{tab_freqSwitch}. The highest rate of changes is observed with the 3~dB cost, although the difference with the case for no switching cost is small and may be caused by chance variations, \textit{e.g.} use of shadowing for DTT. With 6~dB cost, the switching rate is lower with only significant changes in sensed energy having impact on new channel selection.
\begin{table}[htbp]
\centering
\caption{Average number of frequency changes with Bumblebee-based algorithm using different switching costs}
\label{tab_freqSwitch}
\begin{tabular}{|c|c|c|c|}
\hline
Platoon & 0~dB (no cost) & 3~dB cost & 6~dB cost \\
\hline
1 & 3.9 & 4.1 & 3.4 \\
\hline
2 & 4.4 & 4.5 & 3.6 \\
\hline
\end{tabular}
\end{table}

\begin{figure}[htbp]
\centerline{\includegraphics[width=0.5\textwidth]{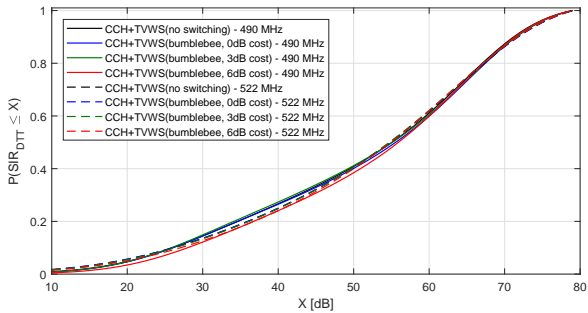}}
\caption{Cumulative distribution of SIR for the protected DTT receivers.}
\label{fig_dttSIR}
\end{figure}

An important aspect of using the TVWS as a secondary system is not to cause the degradation of the DTT service. Therefore, we investigated the Signal-to-Interference (SIR) levels of the DVB-T transmission observed at the locations of the respective DTT receivers as the platoons move across the motorway. According to \cite{DSA_ruler_DTT_2017}, the SIR value of 39.5~dB should be kept in order to provide the required QoS of DTT. Fig.~\ref{fig_dttSIR} presents the empirical cumulative distribution of the SIR values of the primary system obtained in simulations for the protected locations of DVB-T receivers. The results indicate that for every considered strategy, the SIR levels are similar with a slightly better performance achieved using the Bumblebee-based algorithm assuming 6~dB cost. However, for both observed DTT bands, a large number of collected SIR samples is below the required threshold. This indicates that the Bumblebee-based algorithm should be enhanced with the measures to protect the primary users, \textit{e.g.} by modifying the channel switching metric (\ref{switchDecision}). Furthermore, the available spectrum is probably also significantly narrow to effectively mitigate the interference to DTT without any power control applied.

\vspace{-5mm}

\section{Conclusions}
This paper presented a distributed VDSA framework for intra-platoon communications. The Bumblebee-based algorithm for dynamic transmission band selection was adopted to work with platoons of vehicles. The simulation results presented in this work indicated that VDSA has potential to improve the reliability of intra-platoon communications. Proposed approach allowed for making independent transmission channel selection by each platoon. However, in order to maximize the outcomes from VDSA and to sufficiently protect the primary users, more aspects have to be taken into account besides the energy analysis. The energy sensing-based optimization might not be the best tool to estimate the impact of transmissions from other platoons utilizing the same band, as with averaging of the measured energy the probability of collision of two or more packets is not considered. Furthermore, an additional cost related to the interference introduced into the primary system should also be considered in the channel switching metric (\ref{switchDecision}). More sophisticated mechanisms can be developed, where channel switching is complemented with a transmit power control to provide additional measure to minimize the introduced interference. Finally, one can consider also an application of some edge intelligence in form of databases that support the decision-making process of platoons with information on the expected DTT power levels or the transmit power constraints for each band.

\bibliographystyle{IEEEtran}
\bibliography{bibtex.bib}

\end{document}